\documentclass[12pt] {article}
\onecolumn
\usepackage{setspace}

\newtheorem{prop}{Proposition}

\newtheorem{cor}{Corollary}

\newtheorem{lm}{Lemma}

\newtheorem{thm}{Theorem}

\newcommand{\be}{\begin{eqnarray}}
\newcommand{\ee}{\end{eqnarray}}
\newcommand{\benn}{\begin{eqnarray*}}
\newcommand{\eenn}{\end{eqnarray*}}
\def\IR{\rm I \kern-0.20em R}
\newcommand{\utwi}[1]{\mbox{\boldmath $ #1$}}

\newcommand{\bthm}{\begin{thm}}
\newcommand{\ethm}{\end{thm}}

\newcommand{\bcor}{\begin{cor}}
\newcommand{\ecor}{\end{cor}}
\newcommand{\bprop}{\begin{prop}}
\newcommand{\eprop}{\end{prop}}
\newcommand{\blm}{\begin{lm}}
\newcommand{\elm}{\end{lm}}
\newcommand{\beq}{\begin{equation}}
\newcommand{\eeq}{\end{equation}}
\newcommand{\ber}{\begin{eqnarray}}
\newcommand{\eer}{\end{eqnarray}}

\newcommand{\bproof}{\begin{proof}}
\newcommand{\eproof}{\end{proof}}

\newcommand{\bit}{\begin{itemize}}
\newcommand{\eit}{\end{itemize}}
\newcommand{\ben}{\begin{enumerate}}
\newcommand{\een}{\end{enumerate}}
\newcommand{\bdesc}{\begin{description}}
\newcommand{\edesc}{\end{description}}
\newcommand{\beqarrn}{\begin{eqnarray*}}
\newcommand{\eeqarrn}{\end{eqnarray*}}
\newcommand{\bproofof}{\begin{proofof}}
\newcommand{\eproofof}{\end{proofof}}
\newenvironment{rem}{\begin{trivlist}\item[]{\bf
Remark:}\hspace{4mm}}{\end{trivlist}}
\newcommand{\brem}{\begin{rem}}
\newcommand{\erem}{\end{rem}}
\newenvironment{rems}{\begin{trivlist}\item[]{\bf
Remarks}\begin{itemize}}{\end{itemize}\end{trivlist}}
\newcommand{\brems}{\begin{rems}}
\newcommand{\erems}{\end{rems}}
\newtheorem{fact}{Fact}
\newcommand{\bfact}{\begin{fact}}
\newcommand{\efact}{\end{fact}}
\newtheorem{examp}{Example}
\newcommand{\bexamp}{\begin{examp}\rm}
\newcommand{\eexamp}{\end{examp}}
\newtheorem{defn}{Definition}
\newcommand{\bdefn}{\begin{defn}\rm}
\newcommand{\edefn}{\end{defn}}

\newtheorem{alg}{Algorithm}
\newcommand{\balg}{\begin{alg}}
\newcommand{\ealg}{\end{alg}}

\newtheorem{prob}{Problem}
\newcommand{\bprob}{\begin{prob}}
\newcommand{\eprob}{\end{prob}}

\newcommand{\bvtm}{\begin{verbatim}}
\newcommand{\bfig}{\begin{figure}}
\newcommand{\efig}{\end{figure}}
\newcommand{\bcen}{\begin{center}}
\newcommand{\ecen}{\end{center}}

\long\def\comment#1{}

\def \n2{{N_0 \over 2}}

\def \h5{\hspace{0.5in}}

\newcommand{\bs}{{\utwi{s}}}

\newcommand{\bv}{{\utwi{v}}}

\newcommand{\bx}{{\utwi{x}}}
\newcommand{\by}{{\utwi{y}}}

\newcommand{\bF}{{\utwi{F}}}

\newcommand{\bH}{{\utwi{H}}}
\newcommand{\bI}{{\utwi{I}}}

\newcommand{\bT}{{\utwi{T}}}

\newcommand{\bGamma}{{\utwi{\mathnormal\Gamma}}}

\usepackage{epsfig,latexsym,amssymb,amsmath,graphics}

\begin{document}

\title{{\bf Coherent Optical DFT-Spread OFDM}}

\author{Fanggang~Wang, \  \  Xiaodong~Wang\\
Electrical Engineering Dept.\\
Columbia University\\
New York, NY 10027\\
fgwang@inc.cuhk.edu.hk, wangx@ee.columbia.edu}

\date{}
\maketitle

\renewcommand{\baselinestretch}{1.8}
\begin{abstract}
We  consider application of the discrete Fourier transform-spread orthogonal
frequency-division multiplexing (DFT-spread OFDM) technique to
high-speed fiber optic communications. The DFT-spread OFDM is a form
of single-carrier technique that possesses almost all advantages of
the multicarrier OFDM technique (such as high spectral efficiency,
flexible bandwidth allocation, low sampling rate and low-complexity
equalization). In particular, we consider the optical DFT-spread
OFDM system with polarization division multiplexing (PDM) that
employs a tone-by-tone linear minimum mean square error (MMSE) equalizer. We show that such a
system offers a much lower peak-to-average power ratio (PAPR)
performance as well as better bit error rate (BER) performance
compared with the optical OFDM system that employs amplitude
clipping.
\end{abstract}


\section{Introduction}

The high-throughput data transmission over long-haul fiber optic
systems is of considerable current interest. To maximize the
spectral efficiency, polarization multiplexing and coherent
detection have become the key enabling technologies for high-speed
fiber optic communication systems \cite{Ip08}. However,  physical
impairments such as the chromatic dispersion (CD), the polarization
mode dispersion (PMD), and the polarization dependent loss (PDL),
become more severe  as the bandwidth and data rate increase. The
orthogonal frequency-division multiplexing (OFDM) technique has been
widely adopted to cope with the frequency-selective fading of
multipath channels in wireless communications; and it has been
recently introduced to fiber optic systems for high-speed data
transmission \cite{Armstrong:09}. In OFDM systems, the
frequency-domain equalization
 is employed with a single-tap equalizer at each tone, which
significantly reduces the computational complexity compared with the
time-domain equalization in single-carrier systems. However, one
major disadvantage of the OFDM system is the high peak-to-average
power ratio (PAPR). To address this issue, the DFT-spread OFDM has
been developed as an alternative wireless access technique and it
has been adopted as the uplink air interface of 3rd generation partnership project long term evolution (3GPP LTE)
\cite{36300}.

In \cite{For:08} it is observed that the impact of nonlinearity on a
link with periodical dispersion compensation is significantly larger
than that on a link without inline dispersion compensation, which
makes the application of OFDM to the existing infrastructure
questionable. In addition, it is also noted that in a periodic
dispersion map, reducing PAPR at the transmitter might significantly
improve the nonlinear tolerance of the transmission link. In order
to avoid the high cost associated with mitigating the nonlinear
impairments caused by the high PAPR in optical OFDM systems, in this
paper, we consider the coherent optical DFT-spread OFDM to lower
PAPR at the transmitter. In Section 2 we describe the optical
DFT-spread OFDM system that employs polarization division
multiplexing (PDM) and coherent detection. The receiver demodulation
is discussed  in Section 3. In Section 4 we present simulations
results. Section 5 concludes the paper.


\section{The Coherent Optical DFT-spread OFDM System with PDM}
\label{sec_sym}

\subsection{System Descriptions}
A block diagram for the optical DFT-spread OFDM system with PDM is
shown in Fig.~1. We transmit and receive signals on both
polarizations which effectively results in a $2\times 2$
multiple-input multiple-output (MIMO) system.  Consider one of the
two data streams at the transmitter. The bit sequence is first
mapped to the quadrature amplitude modulation (QAM) symbols.  In the
traditional OFDM system, an inverse discrete Fourier transform
(IDFT) is directly applied to the QAM symbols. In the DFT-spread
OFDM system, on the other hand, a DFT is first applied to the QAM
symbols, and then followed by the IDFT operation. After inserting
the cyclic prefix (CP) symbols, the electrical signal is passed
through the digital-to-analog converter  and the low-pass filter,
and  then upconverted to the optical signal. The optical signal
traverses
 in the long-haul fiber link comprising of multiple spans
 to reach the destination. At
the receiver the optical signal is downconverted to the electrical
signal, which is then low-pass filtered and passed through the
analog-to-digital converter. After removing the CP and performing
the DFT on the signals at each polarization, the  two coupled
frequency-domain received signals over $M$ subcarriers are obtained.
We then perform a tone-by-tone MIMO equalization to decouple the two
signal streams. Finally, the IDFT is performed on each decoupled
signal stream to recover the corresponding transmitted QAM symbol
stream.

Optical OFDM system has similar block diagram as Fig. 1 but without the DFT or IDFT modules marked as dark blocks. Intuitively, the OFDM system has a high PAPR since after IDFT at the
transmitter multiple input QAM symbols could be phase-synchronously
added together and therefore causes high signal amplitude (peak
power).  The high PAPR of the signal   decreases the system power
efficiency and makes the transmission more susceptible to nonlinear
impairments of the fiber link. On the other hand, the DFT and IDFT cancel each other and thus the DFT-spread OFDM system is essentially a single-carrier technique which in general has much a lower PAPR. However, note that the
optical DFT-spread OFDM considered here is fundamentally
 different from the traditional  single-carrier scheme \cite{Shieh08},
 in that the former exhibits the advantages of both the single-carrier
 system (i.e., low PAPR) and the OFDM system (i.e., tone-by-tone single
 tap equalizer).

\subsection{MIMO Channel Model for PDM}
For the long-haul optical fiber transmission, the fiber link
comprises $n_E$ fiber spans. We consider a periodic dispersion map
on existing 10Gb/s infrastructure. In this case, the high PAPR
caused by
 dispersion is trivial and we therefore  focus on reducing the PAPR at the
transmitter.
We consider three typical  distortions in the fiber channel, CD,
first-order PMD, and PDL. The $2 \times 2$ transfer function at
  subcarrier $m$ corresponding to a DFT-spread OFDM symbol
is given by \cite{Shieh08}
\begin{equation}
\bH_m    = e^{j\phi} e^{j\Phi _D (f_m )} \bT_m, \label{eq1}
\end{equation}
where  $\phi$ is  a common phase error (CPE) noise owing to the
phase noises that varies for different OFDM symbols.  In (\ref{eq1})
$\Phi_{D}(f_m)$ is the phase dispersion owing to CD and given  by
\begin{equation}
\Phi_D (f_m) = \frac{\pi c D L f_m^2}{f_c^2},\quad \text{with}\ f_m
= \frac{m-1}{t_s}.
\end{equation}
where $t_s$ is the actual time duration of one DFT-spread OFDM
symbol not including the cyclic prefix; $c$ is the speed of the
light; $D$ is the CD parameter in unit of  ps/pm/km; $L$ is the
total length of the multi-span fiber link; and $f_c$ is the carrier
frequency of the laser. Here we assume that the periodic dispersion
map is employed and the CD is completely compensated for  at each
span. Hence $\bT_m$ in (\ref{eq1}) is the Jones matrix with
the dispersion compensation fiber (DCF)
inserted at each stage, given by
\begin{equation}  \label{formu_mueller}
\bT_m = \prod \limits_{l=1}^{n_E} e^{- j \frac {\Phi _D (f_m
)}{n_E}} \left[ {\begin{array}{*{20}c}
   1 & 0  \\
   0 & k_l  \\
\end{array}} \right ]\cdot \left[ {\begin{array}{*{20}c}
   e^{j\pi(f_c+f_m)\tau_l} & 0  \\
   0 & e^{-j\pi(f_c+f_m)\tau_l}  \\
\end{array}} \right]\cdot
\left[ {\begin{array}{*{20}c}
   \cos\theta_l & \sin\theta_l  \\
   -\sin\theta_l & \cos\theta_l  \\
\end{array}} \right],
\end{equation}
where $k_l$ denotes the attenuation factor of PDL; $\tau_l$ is the
differential group delay (DGD); $\theta_l$ is the uniformly random
rotation angle \cite{Hau09}. For each span, the DGD is a random
variable following the Maxwellian distribution. However, in the same
span, the DGDs of different subcarriers are the same.

\section{The Linear MMSE Coherent Receiver}

Let $\bF$ be the $M\times M$ DFT matrix with its element given by
\begin{equation}
  F_{\alpha,\beta}  = (1/\sqrt M )e^{ - j2\pi (\alpha - 1)(\beta -
1)/M}.
\end{equation}
Denote the transmitted QAM  symbols on all subcarriers along
the $k$th polarization as $\bs^{(k)} \triangleq [s_1^{(k)}
,s_2^{(k)}, \cdots ,s_M^{(k)} ]^T$. As shown in Fig.~1, a DFT
operation is first applied to $\bs^{(k)}$, to obtain
 \be \bx^{(k)} \triangleq [x_1^{(k)} ,x_2^{(k)} ,
\cdots ,x_M^{(k)} ]^T =\bF\bs^{(k)},\ee which is effectively the
input to a traditional OFDM system. At the receiver, we assume the
frequency and time offset can be perfectly estimated and compensated
before detection.
    Then the received  signal on the $m$th
subcarrier is given by
\begin{equation}
\label{formu_sys} \by_m  = \bH_m \bx_m  + \bv_m,\quad m = 1, \ldots
,M,
\end{equation}
where $\bx_m \triangleq [x_m^{(1)} ,x_m^{(2)} ]^T$, and $\bv_m\sim
\mathcal{N}_c(\mathbf{0},\sigma^2\bI)$.

To demodulate the symbol vector $\bs^{(k)}$, we first estimate the
DFT-spread symbol $\bx^{(k)}$   and then recover the data symbol
vector $\bs^{(k)}$ by an IDFT. In particular, a linear MMSE estimate
of $\bx_m$ based on $\by_m$ in (\ref{formu_sys}) is given by
\cite{Prasad:08}
\be
 \hat \bx_m &=& \bGamma \bH_m^{\dag} (\sigma^2\bI +
\bH_m \bH_m^{\dag} )^{ - 1} \by_m \nonumber \\
&=& \bGamma (\sigma^2\bI + \bH_m^{\dag} \bH_m )^{ - 1} \bH_m^{\dag}
\by_m , \ \ m=1, \ldots, M, \label{eq2} \ \quad  \ee where
$(\cdot)^{\dag}$ denotes conjugate transpose, and $\bGamma$ is a
diagonal matrix with $\gamma_{k,k} = 1/\Big(1 - [(\bI + \bH_m
\bH_m^{\dag}/\sigma^2 )^{ - 1}]_{k,k}\Big)$. Note that the
tone-by-tone linear MMSE equalization in (\ref{eq2}) involves
inverting $2 \times 2$ matrices and hence the computational
complexity is not significant. Denote $\hat \bx^{(k)} \triangleq
[\hat x_1^{(k)} , \ldots ,\hat x_M^{(k)} ]^T $. We finally  apply
the IDFT on $\hat \bx^{(k)}$ to obtain the estimated data symbols
 \be \hat \bs^{(k)}= \bF^{\dag} \hat \bx_{}^{(k)},   \ \ k=1,2. \ee


\section{Simulation Results}

In this section we provide simulation results to compare the optical
DFT-spread OFDM system with the optical OFDM system in terms of both
the PAPR performance and the bit error rate (BER) performance. The
number of  subcarriers is  $M=256$.   A long-haul fiber optic system
is considered with $n_E=12$ cascaded spans (each of length $L=80
km$). The laser wavelength is $\lambda=1.55\mu m$. The CD parameter
is $D = 17 ps/nm/km$. We assume that periodic dispersion map is
employed and the CD is compensated for by DCF after each span. The
DGD parameter is $D_p = 0.15 ps/\sqrt{km}$. The mean value of DGD is
$\sqrt{8/(3\pi)}D_p \sqrt{n_E L}$. A typical PDL is $0.1$dB, where
$\text{PDL[dB]}=-20\log(k_l)$.  We consider both QPSK and 16-QAM
modulations. The data rate is 25G symbols/s, i.e.,  100Gb/s for QPSK
and  200Gb/s for 16QAM.

Figure~2 illustrates the PAPR performance of the two  systems for
QPSK and 16-QAM. It is seen that the OFDM system exhibits a much
higher PAPR than the DFT-spread OFDM system. In general 16-QAM has a
higher PAPR than QPSK. However, their PAPRs are similar in the OFDM
system. Amplitude clipping is a typical method to lower the PAPR in
OFDM systems. We used a clipping ratio (CR) of 3dB, defined as
$\mbox{CR}=20\log_{10}(A/P)$, where $P$ is the power of transmitted
signal, and $A$ is the maximum transmitted signal magnitude. It is
seen that clipping can indeed significantly
 reduce the PAPR in the OFDM system; however, the PAPR in the clipped
 OFDM system is still much larger than that of the
 DFT-spread OFDM system. On the other hand,   clipping substantially degrades the
 bit error rate (BER)
performance. As shown in Fig.~3, the DFT-spread OFDM and the OFDM
systems have the same  BER performance. But after the 3dB clipping
on the OFDM signal,  a  0.8dB loss is incurred at the BER of
$10^{-3}$ for QPSK, and for 16QAM the loss   due to clipping is
about 8dB. In summary, compared with the traditional optical OFDM
system that employs amplitude clipping, the DFT-spread OFDM system
offers both a lower PAPR and a better BER performance.

\noindent {\it Discussions:} \ Note that the effect of nonlinearity
depends on the instantaneous power of the signal according to the
 Schrodinger's equation \cite{Agra}. It is shown in Fig.~2 that the PAPR
performance of the DFT-spread OFDM signal is better than both the
clipped and unclipped OFDM signals. This means that for a given
average transmit power, the peak power of both the clipped and
unclipped OFDM signals will be larger than that of the DFT-spread
OFDM signals; and hence they are more susceptible to nonlinear
distortion than the DFT-spread OFDM signals. On the other hand,
various effective nonlinear impairment compensation methods have
been proposed \cite{Mateo}. Hence one can envision that if these
techniques are applied to the system, only linear channel distortion
needs to be considered, under which we have shown the DFT-spread
OFDM offers better BER performance than the clipped OFDM.

\section{Conclusions}
We have considered the optical DFT-spread OFDM system with
polarization division multiplexing and coherent detection.
 Compared with the conventional single-carrier systems,
the DFT-spread OFDM system   has the advantages of flexible
bandwidth allocation, high spectral efficiency and low sampling
rate, and low-complexity equalization. Compared with the optical
OFDM system with amplitude clipping, the DFT-spread OFDM
 system offers both better BER performance  and
 a much  lower PAPR, with little attendant increase in
 the transceiver complexity.


\newpage
\begin{figure*}[t]
\centering\includegraphics[width=15cm]{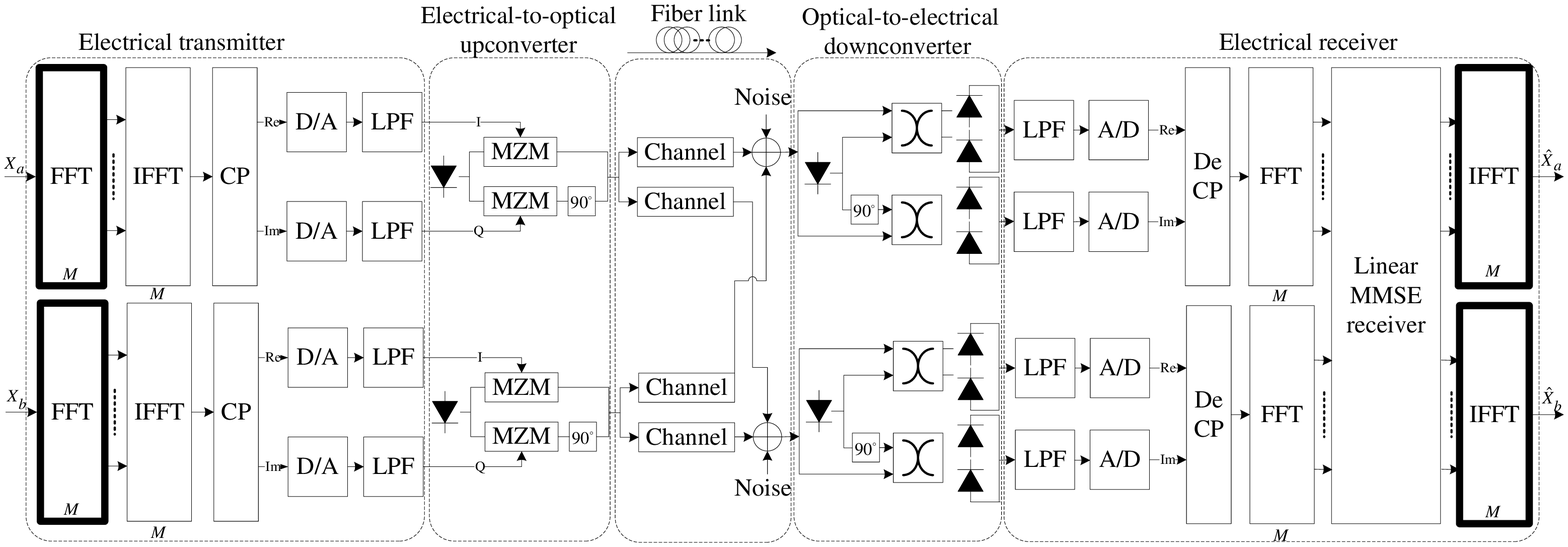} \caption{Block
diagram for an optical  DFT-spread OFDM system with PDM and coherent
receiver.}
\end{figure*}

\begin{figure}
\centering\includegraphics[width=5in]{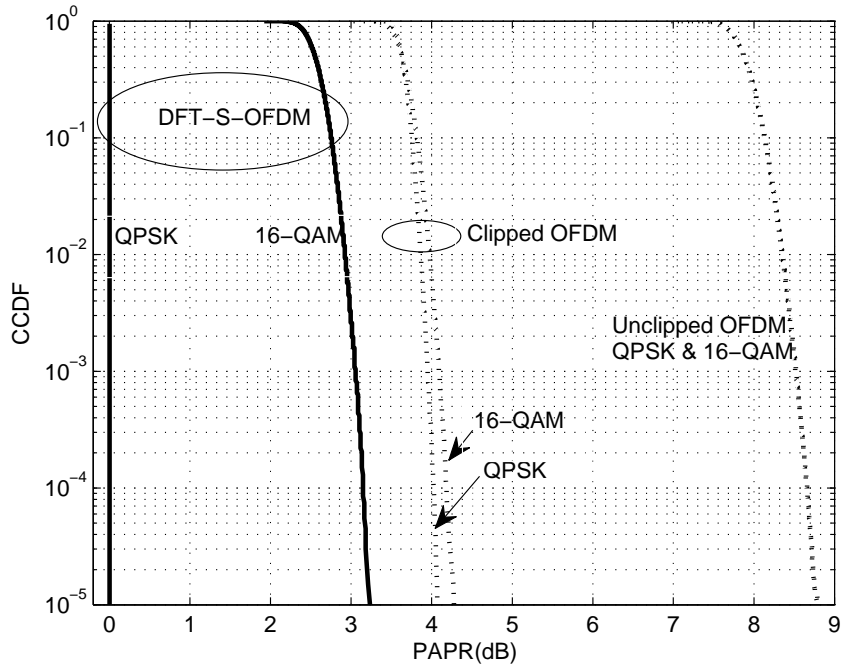} \caption{Comparison
of the complementary cumulative distribution function (CCDF) of the
PAPR.} \label{papr}
\end{figure}

\begin{figure}
\centering\includegraphics[width=5in]{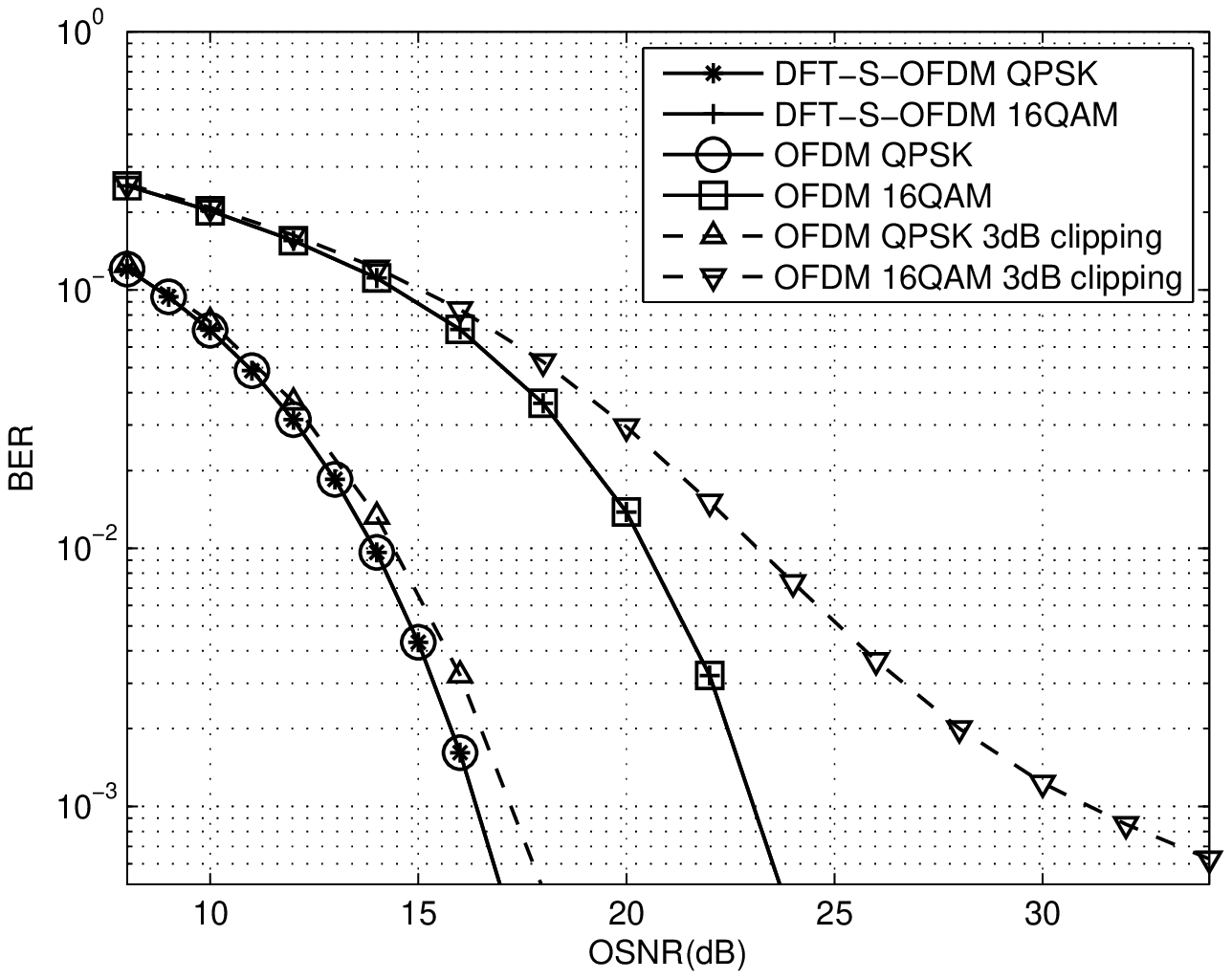} \caption{BER
comparison.} \label{ber}
\end{figure}


\begin{thebibliography}{99}

\bibitem{Ip08}
E.~Ip,  A.~Lau,  D.~Barros,  and  J.~Kahn,
\newblock Coherent detection in optical fiber systems.
\newblock {\em Opt. Express}, vol. 16, no. 2, pp. 753--791, 2008.




\bibitem{Armstrong:09}
J.~Armstrong.
\newblock OFDM for optical communications.
\newblock {\em J. Lightwave Technol.}, vol. 27, no. 3, pp. 189--204, 2009.


\bibitem{36300}
ETSI Std.
\newblock Evolved universal terrestrial radio access (E-UTRA) and
evolved universal terrestrial radio access network (E-UTRAN);
overall description; stage 2.
\newblock 3GPP TS36.300, Rev. 9.1.0, 2009.

\bibitem{For:08}
K.~Forozesh,  S.~L.~Jansen, and S.~Randel, etc.
\newblock The influence of the dispersion map in coherent optical
OFDM transmission systems.
\newblock {\em IEEE/LEOS Summer Topical Meetings, 2008 Digest of
the}, pp. 135-136, Jul. 2008.



\bibitem{Shieh08}
W.~Shieh, X.~Yi, Y.~Ma, and Q.~Yang.
\newblock Coherent optical OFDM: has its time come?
\newblock {\em J. Opt. Netw.}, vol. 7, no. 3, pp. 234--255, 2008.


\bibitem{Hau09}
F.~Hauske, M.~Kuschnerov, B.~Spinnler, and B.~Lankl.
\newblock Optical performance monitoring in digital coherent receivers.
\newblock {\em J. Lightwave Tech.}, vol. 27, no. 16, pp. 3623--3631, 2009.


\bibitem{Prasad:08}
N.~Prasad, S.~Wang, and X.~Wang.
\newblock Efficient receiver algorithms for DFT-spread OFDM systems.
\newblock {\em IEEE Trans. Wireless Commun.},  vol. 57, no. 7, pp. 2797-2808, 2009.


\bibitem{Agra} G.P. Agrawal. {\em Nonlinear Fiber Optics}, 3rd Ed. Academic
Press,   San Diego,  2001.

\bibitem{Mateo} E.~Mateo,  L.~Zhu, and G.~Li.   Impact of XPM and FWM on the
digital implementation of impairment compensation for WDM
transmission using backward propagation. {\em  Optics Express}. vol.
16, no. 20, pp. 16124-16137, 2008.


\end{thebibliography}
\end{document}